\documentclass[draft]{agujournal}

\journalname{}

\begin{document}

\title{Microphysical effects of water content and temperature on the triboelectrification of volcanic ash on long timescales}

\authors{Joshua M\'endez Harper\affil{1},  Leah Courtland\affil{2}, Josef Dufek \affil{3}, and Julian McAdams\affil{3}}

\affiliation{1}{Department of Physics, Emory University}
\affiliation{2}{Physics and Earth-Space Science, University of Indianapolis}
\affiliation{3}{Department of Earth Sciences, University of Oregon}

\correspondingauthor{Joshua M\'endez Harper}{joshua.mendez@emory.edu}

\begin{keypoints}
\item The triboelectrification of ash in low-energy collisions is modulated by humidity
and temperature on long timescales
\item Triboelectrification may be effective in the gas-thrust region, but not higher in the plume
\item The reduction in charging efficiency under humid environments could explain hiatuses in electrical activity at some volcanoes
\end{keypoints}

\begin{abstract}

The effects of water and temperature on the triboelectrification of granular materials have been reported by numerous authors, but have not been studied robustly in the context of volcanic plumes. Here, we present the results of a set of experiments designed to elucidate how environmental conditions modulate the triboelectric characteristics of volcanic ash in the upper region of the convective column. We find that that small amounts of water can reduce the charge collected by micron-sized ash grains by up to an order of magnitude. Increasing temperature at a constant relative humidity also appears to decrease the amount of charge gained by particles. Analysis of our data shows that if particles undergo low-energy, low-frequency collisions in humid environments under long timescales, charge dissipation dominates over charge accumulation. Thus, our work suggests that triboelectric charging may be an inefficient electrification mechanism outside of the gas-thrust region where collision energies and rates are high and residence times are low. 

\end{abstract}

\section{Introduction}

Volcanic granular flows can become highly electrified as evidenced by changes in the ambient electric field beneath plumes \citep{nagata1946charge, hatakeyama_variation_1947, hatakeyama_disturbance_1951, lane1992electric, miura_atmospheric_1996}, elevated charge densities detected on settling ash particles \citep{gilbert_charge_1991, miura_atmospheric_1996, harrison2010self}, observations of tephra aggregates bound together by electrostatic forces  \citep{james_experimental_2002, james2003density, telling2013ash}, attenuation of electromagnetic radiation propagating through plumes \citep{mendez2019gps}, and spectacular displays of lightning \citep{thomas_electrical_2007, mcnutt_volcanic_2010, behnke_observations_2013, behnke_changes_2015, aizawa_physical_2016, cimarelli_multiparametric_2016, behnke_investigating_2018, mendez2018inferring}. Field observations and experimental efforts suggest that electrostatic processes in volcanic granular flows (a) arise from a multitude of charging and charge-separation mechanisms operating at different locations and times within the plume \citep{james_volcanic_2000, thomas_electrical_2007, mcnutt_volcanic_2010, behnke_observations_2013, behnke_changes_2015, mendez_harper_electrification_2015, mendez_harper_effects_2016}; (b) reflect eruption kinematics \citep{behnke_observations_2013, behnke_changes_2015, cimarelli_multiparametric_2016, mendez2018inferring}; and (c) correlate with environmental conditions in the plume \citep{mather2006electrification, arason2011charge, nicora2013actividad, vaneaton2016volcanic}. 

\begin{figure}[h]
	\centering
	\includegraphics[width=0.8\textwidth]{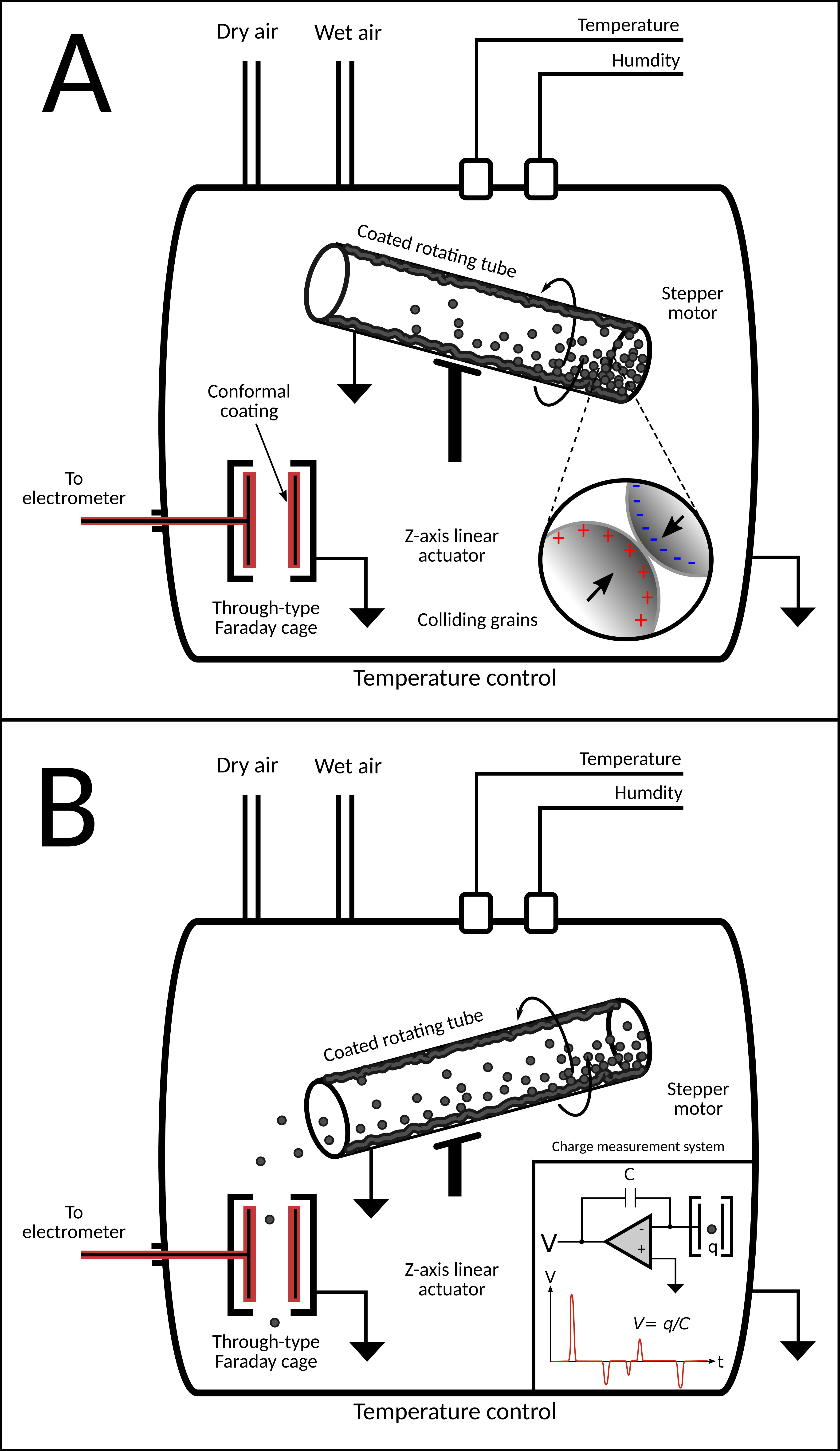}
	\caption{Schematic of laboratory apparatus, adapted from \citet{mendez_harper_electrification_2017}. Charging device consists of a hollow aluminum tube open at one end and connected to the grounded shaft of a stepper motor at the other. A. The open end of the tube is raised by a second stepper motor to enable particles to be input to the device. The interior of the tumbler is coated with particles of identical size and composition as the input particles such that tribocharging arises from particle-particle collisions only. B. The tube is shifted downward, still rotating, causing particles to fall out of the tube and into the Faraday cage where the charge on individual particles is measured.}
	\label{figure1}
\end{figure}  

A number of electrification mechanisms have been invoked to explain electrostatic processes in jets and plumes, including fragmentation charging \citep{james_volcanic_2000, mendez_harper_electrification_2015}, triboelectric charging \citep{houghton2013tribo, cimarelli_experimental_2014, mendez_harper_effects_2016, mendez_harper_electrification_2017}, charging arising from the decay of radioactive compounds in ash \citep{houghton2013tribo}, and mechanisms comparable to those thought to operate in thunderclouds \citep{mather2006electrification, nicora2013actividad}. \textit{Triboelectrification}--referring collectively to the electrification arising from frictional or collisional interactions between two or more surfaces--has been recognized since antiquity \citep{iversen2012life} and has been detected or inferred to exist in many of natural granular systems \citep{farrell2004electric, dou2017electromagnetic, fabian2001measurements, forward_particle-size_2009, harrison2016applications, sternovsky2002contact, forward2009dtriboelectric, mendez_harper_electrification_2017, helling2013ionization, hodosan2016lightning, mendez2018tribo}. Carefully controlled experimental studies have shown that volcanic ash readily charges triboelectrically \citep{hatakeyama_disturbance_1951, james_volcanic_2000, houghton2013tribo, mendez_harper_effects_2016, mendez_harper_electrification_2017}. Because pyroclast-pyroclast collisions occur within the plume throughout all stages of the eruption, triboelectrification may contribute to electrical processes in volcanic granular systems.

Despite being one of the oldest manifestations of electricity known to humans, triboelectric charging involves underlying physical processes that remain inadequately understood \citep{lowell1986triboelectrification, lacks_effect_2007, lacks_contact_2011, shinbrot2014granular}. For instance, experimental efforts have yet to unequivocally reveal the identities of the charge carriers being exchanged during collisions. Some investigators suggest that triboelectric charging arises from the exchange of electrons trapped in unfavorable energy states \citep{lowell1980contact, lowell1986triboelectrification, lacks_effect_2007, duff2008particle, forward_charge_2009, kok2009electrification}, whereas other works point to an ionic process \citep{gu2013role, zhang2015electric, xie_effect_2016}. Others still have argued that triboelectric charging requires the transfer of material fragments from one surface to another (see, for example, \citet{salaneck1976double}), implying that triboelectricity is nominally a manifestation of fragmentation charging. Given the wide gaps in knowledge surrounding triboelectric and fractoelectric processes, for the time being we will consider them two separate processes. 

Another aspect of triboelectric charging that remains poorly quantified is the role of ambient conditions under which electrification takes place --specifically, the presence of volatiles. While a number of investigators have attempted to elucidate how water influences triboelectric behavior, the results of these studies sketch an inconclusive, sometimes contradictory, picture. Some studies report that charging decreases monotonically with increasing water content, citing either an increase in air conductivity or the growth of conductive water layers (through wetting) on the charged surfaces \citep{harper1957generation, tada1995direct, greason_investigation_2000, diaz2004semi, burgo2011electric, schella2017influence, yao2014experimental}. Others report that low amounts of water actually \textit{enhance} frictional electrification because H\textsuperscript{+} and OH\textsuperscript{-} tend to segregate on large and small particles, respectively, producing the oft-cited size-dependent bi-polar charging \citep{hiratuska2012effects, gu2013role, xie_effect_2016}.  While no consensus has been reached, the role of water in triboelectric charging is becoming increasing difficult to dismiss. Likewise, experimental work suggests that temperature has an effect on frictional charging processes. For example, \citet{greason_investigation_2000} found that the magnitude of charge on beads in a vibrated bed decreases with increasing temperature. Additionally, \citet{olsen2018schottky} notes that some researchers (including \citet{wen2014applicability}, \citet{su2015low}, and \citet{lu2017temperature}) report a power output decrease with temperature in of triboelectric nano-generators. Temperature variations may also affect the presence of contaminants on surfaces, changing their ability to gain or retain charge \citep{lowell1980contact}. For instance, contaminants may change the effective work function of a surface, which, in turn, influences the direction and magnitude of charge exchanged when it interacts with other surfaces. Given the wide range of effects temperature and humidity appear to have on the triboelectrification of granular materials, we suspect such effects can be extended to volcanic systems.  Erupted particles may have temperatures ranging from several hundreds of degrees near the vent to temperatures below freezing in the elevated, distal plume. Similarly, depending on the character of an eruption, ash particles may undergo collisions in relatively dry environments or be surrounded by humid air.

In some regards, water and temperature have already been recognized as factors that tune electrical phenomena in plumes. As mentioned above, studies indicate that electrical discharges in fully developed plumes involve electrification mechanisms similar to those found in thunderclouds \citep{mather2006electrification, arason2011charge,nicora2013actividad, vaneaton2016volcanic, hargie2018globally}. When ejected material reaches sufficiently high elevations, charging may be primarily driven by collisions between ice-coated particles and graupel (wet hail). Through experiments, \citet{durant_ice_2008} found that ice starts to nucleate on ash grains when the ambient temperature falls in the range of -10 to -20\textsuperscript{o} C. Yet, there is evidence, although sparse, that water in solid form need not be present for environmental conditions to modulate the electrification capacity of ash. \citet{james_volcanic_2000}, for instance, reported that particles electrified via fragmentation and frictional processes were possibly influenced by relative humidity (RH), but those authors indicate that their results may reflect anomalous behaviors of their measurement equipment rather than changes in the amount of charge collected by the ash. More recently, \citet{stern_temp_2019} used a laboratory-scale shock tube to approximate the eruption of a wet (i.e. completely saturated), hot granular flow. These investigators found that the number of spark discharges decreases with increasing water content (from 0 - 27 wt. \%) in a 300\textsuperscript{o} C jet. \citet{stern_temp_2019} also explored the role of temperature within the range of 25 - 320\textsuperscript{o} C and report a moderate increase in electrical activity with increasing temperature. The authors hypothesize that this increase was likely due to changes in jet dynamics, not to intrinsic changes in the ash's ability to collect charge. We note that these experiments involved high-energy, possibly disruptive, particle-particle collisions and the observed electrical discharges may have been from both triboelectric and fragmentation charging.  

Here, we present results of a new set of experiments designed to elucidate the response of triboelectric charging to changing atmospheric water content and temperature. Our experiments are relevant to maturing plumes in which ice has \textit{not} yet nucleated onto particles, but in which particles have been lofted sufficiently long for liquid water films to form on their surfaces. Specifically, we explore the region of the volcanic system in which particles are being carried by a weakening convective column and particle-particle collisions involve low energies. Thus, our experiments were designed to reduce contributions from other electrification mechanisms such as fragmentation charging. In agreement with other work, we find that increasing water content diminishes the capacity of ash to hold charge. However, we observe that the amount of water required to severely hinder the efficacy of triboelectric charging is much smaller than previously reported by \citet{stern_temp_2019}.  We propose that minute amounts of water in volcanic granular flows (either magmatic or entrained) may effectively quench electrification through frictional, non-disruptive interactions if the residence times of the ash are high. Thus, our experiments suggest that triboelectric charging may be efficient only near the vent region of the volcanic jet where particles are being generated and new surfaces have not yet had time to absorb water from the environment.  

\section{Methods}

To investigate the ability of ash particles to charge through frictional and collisional interactions under a variety of environmental conditions, we adapted the apparatus described in \citet{mendez_harper_electrification_2017} (shown schematically in figure \ref{figure1}), which can characterize the charge on single micron-sized ash grains. The device consists of a hollow aluminum tube open at one end and connected to the grounded shaft of a stepper motor at the other. The open end of the tube can be raised or lowered by a second stepper motor. At the onset of an experiment, the open end of the tube was raised 5 degrees from horizontal and 100 mL of particles were placed in its interior. The interior of the tumbler was coated with particles of identical size and composition as the free grains such that tribocharging arises from particle-particle collisions only. To neutralize any charge that may have been transferred to the particles during handling, the sample in the tube was sprayed with a bipolar ionizing air gun for 1 minute. The apparatus was then placed within a chamber in which we could control relative humidity (RH) and temperature. We performed two sets of experiments: 1) Varying humidity at a constant temperature and 2) varying temperature at a constant humidity. The environmental conditions we explored are summarized in table \ref{tb1}. Note that both variables ultimately control the amount of water present in the experiment.  

\begin{table}   
	\caption{Humidity and temperature conditions explored in the experiments. The table also includes equivalent total water content ranges for set conditions.} 
	\begin{center} 
		\begin{tabular}{cccccc} 
			Experiment &  Temperature & RH & Total water content\\ 
			\hline 
			Variable RH & 20\textsuperscript{o} C  & 0 -- 50\% & 0 -- 0.012 kg m\textsuperscript{-3} \\
			Variable T & -20 -- 40\textsuperscript{o} & 30\% & 0 -- 0.015 kg m\textsuperscript{-3}\\
		\end{tabular} 
	\end{center} 
	\label{tb1}
\end{table}

For a given combination of temperature and relative humidity, we allowed the conditions in the chamber to stabilize (as measured by the output of humidity and temperature sensors). We then waited for an hour before starting an experiment. Once equilibrated, particles were charged by running the tumbler. Particles exchanged charge as they collided with each other and with the coated interior walls of the tube. The tube was spun at a rotation rate of 0.5 m/s. For these dynamical conditions, \citet{mendez_harper_electrification_2017} found that particles achieved electrostatic steady-state in approximately 15 minutes. Thus, each experiments lasted 20 minutes. In contrast to the energetic granular dynamics in the experiments of \citet{stern_temp_2019}, the collision energies produced in the tumbler are more representative of conditions in convective columns transitioning to mature plumes than those in the gas-thrust region. The low energy interactions between particles did not produce obvious mechanical modification to the grains (as quantified by microscopy analysis performed before and after an experiment). Thus, we make the assumption that contributions from fragmentation charging could be neglected.

At the end of the charging period, the rotating tube was inclined downward, causing the particles to roll out under gravity. Upon leaving the tube, the grains passed through an open-ended Faraday cage connected to a charge amplifier similar to that described by \citet{watanabe2007new}. Such setup allowed us to measure charge in a non-contact manner. To minimize problems associated with increased leakage currents caused by raising humidity (as reported in \citet{james_volcanic_2000}), we encased the Faraday cage in a high-impedance acrylic conformal coating (see figure \ref{figure1}). The charge amplifier was maintained outside of the environmental chamber to prevent drift in the electronics due to temperature changes.

The charge, $q$, on a particles falling through the sensing volume of the Faraday cup produces a voltage pulse at the output of the charge amplifier: 

\begin{equation} \label{eq1}
V_o = \frac{-q}{C_f}.
\end{equation} 

In equation \ref{eq1}, C\textsubscript{f} is the feedback capacitance of the charge amplifier (in this case 100 pF). Our measurement system can measure charges on individual particles as low as 1 fC (or 10\textsuperscript{-15} coulombs), enabling characterization of the magnitude and polarity of charge on individual ash particles. More details can be found in the supplemental document of \citet{mendez_harper_electrification_2017}.

To measure the water content in the chamber we used a Honeywell HIH-4030 series humidity sensor. This device has a nearly linear response with humidity. Additionally, the sensor has an operating temperature range of -40 to 85 \textsuperscript{o}C and allows the user to compensate for temperature induced error. Temperature was measured using a K-type thermocouple and a Maxim MAX31855 thermocouple amplifier with a cold-junction reference. 

In these experiments, we employed ash from volc\'an Popocat\'epetl (``Smoking mountain'' in N\'ahuatl, the lingua franca of pre-Colonial Mesoam\'erica), State of Puebla, M\'exico. The material was washed and sieved so as to have a nominal particle size distribution between 125-250 $\mu$m. We used this size distribution because smaller particles tended to adhere to the inside of the tumbler once the charging period was concluded. The material was washed with deionized water prior to running experiments to reduce the effects of contaminants and small fragments adhered to the surfaces of the particles. To rid the material of any residual moisture from the cleansing process, we baked the ash at 200 \textsuperscript{o}C for 24 hours immediately prior to conducting our experiments. We note that these preparation steps were necessary to produce homogeneous samples across all experiments. 

\begin{figure}[h]
	\centering
	\includegraphics[width=\textwidth]{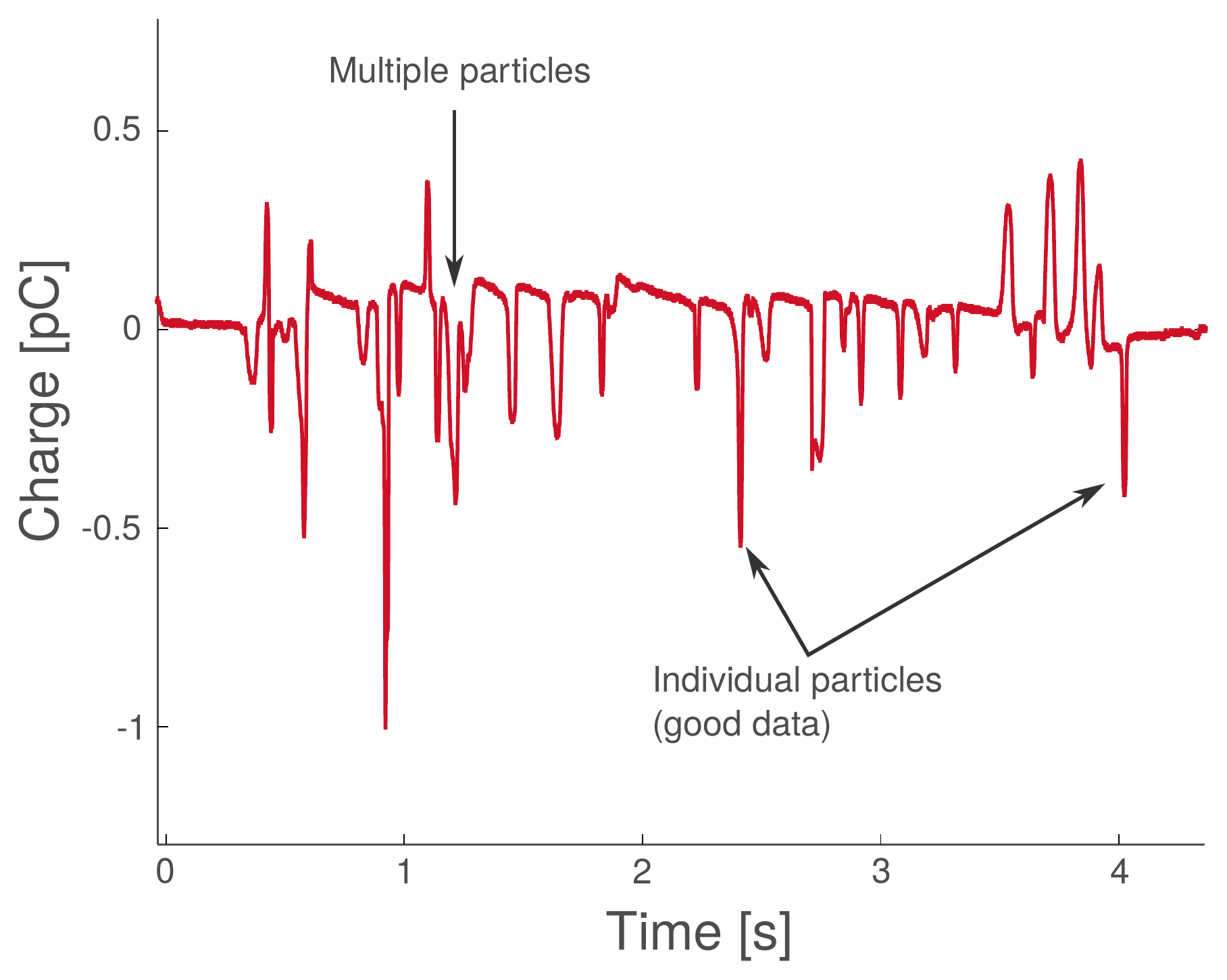}
	\caption{Example of data collected from the charge amplifier as particles fall through the Faraday cage. Peaks represent the magnitudes of charge on individual particles. Overlapping peaks or very broad peaks are representative of multiple particles passing through the sensing volume at the same time. We exclude these peaks from further analysis. Note that the majority of peaks are negative, resulting from the fact that free particles have areas which are smaller then the collective surface area of particle adhered to the tube. Experiment shown was conducted at 20 \textsuperscript{o}C and 30\% RH.}
	\label{figure2}
\end{figure}  

\section{Results}

An example of raw data collected from the output of the charge amplifier during an experiment conducted at 20 \textsuperscript{o}C and 30\% RH (corresponding to a total water content of $\sim$ 0.005 kg m\textsuperscript{-3}) is provided in figure \ref{figure2}. Each pulse represents a charged particle passing through the Faraday cage. Very broad pulses or or overlapping pulses denote conditions in which more than one particle traversed the sensing volume at a time. We exclude such data from further analysis. In agreement with size-dependent bipolar charging (by which small particles gain negative charge whereas large particles acquire positive charge), most particles in our experiments acquired negative charge given that their surface areas are much smaller than the aggregate, interior area of the coated tumbler \citep{forward_particle-size_2009, mendez_harper_electrification_2017}. On average, grains obtained net charges on the order of 10\textsuperscript{-13}  C - 10\textsuperscript{-12} C for both sets of experiments.

\begin{figure}[h]
	\centering
	\includegraphics[width=0.8\textwidth]{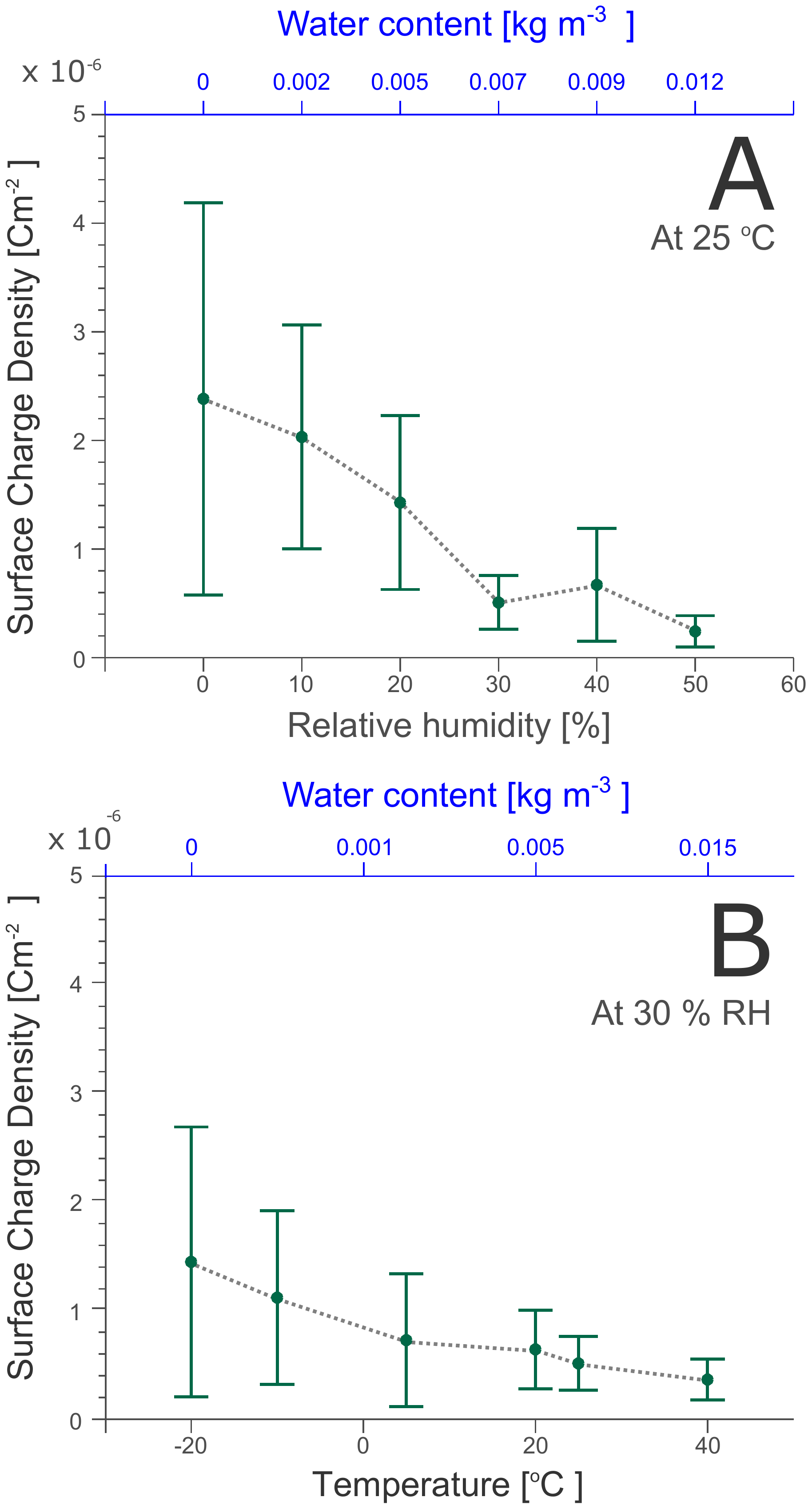}
		\caption{A. Mean charge densities for experiments conducted at 25 \textsuperscript{o}C and relative humidity varying from 0 to 50\%. B. Mean charge densities for experiments conducted at 30\% RH and temperatures ranging from -20 to 40 \textsuperscript{o}C. Each data point represents the agglomeration of $\sim$100 individual particle measurements.}
	\label{figure3}
\end{figure}

A common way to quantify electrification in a granular substance is to compute the material's \textit{surface charge density} or the charge normalized by the particle's surface area: $\sigma$ = $q$/($\pi D$\textsuperscript{2}). For brevity, we use the nominal mean, spherical-equivalent diameter of the ash particles for all computations in this work (i.e.  $D \approx$ 188  $\mu m$). The mean charge densities observed in our experiments are on the order of 10\textsuperscript{-6} C m\textsuperscript{-2}. The relationships between temperature and relative humidity and charge density are rendered in figure \ref{figure3}a and figure \ref{figure3}b, respectively. On those same plots we also report the total water content, in units of kg m\textsuperscript{-3}, for the given combinations of temperature and relative humidity conditions (see upper axes). The error bars represent the span of the data. Each data point represents the agglomeration of $\sim$100 individual particle measurements. 

Our experiments show that both temperature and relative humidity have important effects on the magnitude of charge density acquired by ash particles through collisional interactions. In agreement with other work, the degree of charging decreases monotonically with increasing relative humidity \citep{greason_investigation_2000, schella2017influence, kolehmainen2017effect}. However, there appears to be a sharper decrease in the magnitude of charge density between 20-30 $\%$ RH. Overall, we find that a 50\% increase in relative humidity at 25 \textsuperscript{o}C causes nearly an order of magnitude drop in the mean charge density. We also observe that charging of the volcanic sample decreases as the temperature increases. As discussed above, similar behaviors have been noted in non-volcanic materials by a number of investigators (e.g. \citet{greason_investigation_2000},\citet{su2015low}, \citet{lu2017temperature}, and \citet{olsen2018schottky}).

\section{Discussion}

Our data can be analyzed through the Greason equation, which describes the rate of triboelectric charging in a granular material \citep{greason_investigation_2000}:

\begin{equation} 
\frac{dq(t)}{dt} = \alpha [q_s - q(t)] - \beta q(t)
\label{eq2}
\end{equation} 

Above, $q(t)$ is the charge on an ash particle at time $t$, $q_s$ is the maximum theoretical charge that can be sustained on a particle before the surrounding gas undergoes breakdown (for air, at 1 bar, the value of $q_s$  = 2.66 $\times$ 10\textsuperscript{-5} $\pi D^2$  coulombs is typically used), $\alpha$ is a constant proportional to collision rate and the efficiency of charge exchange during a collision, and $\beta$ encompasses charge loss mechanisms. Integrating the Greason equation, leads to:

\begin{equation} 
q(t) = q_o e^{-(\alpha + \beta)t} + q_s \frac{1}{1+\beta/\alpha} [1 - e^{-(\alpha + \beta)t}]
\label{eq3}
\end{equation}

Above, $q$\textsubscript{o} is any initial charge on grains prior to the triboelectric process (for instance, charge gained through fractoelectric charging).  If $t \rightarrow \infty$ and $q$\textsubscript{o} = 0, equation \ref{eq3} reduces to:

\begin{equation} \label{eq4}
q_{ss} = \frac{2.66 \times 10\textsuperscript{-5} \pi D\textsuperscript{2}}{1+\beta/\alpha}
\end{equation} 

where $q$\textsubscript{ss} is the steady-state charge. Alternatively, equation \ref{eq4} can be expressed in terms of charge density by dividing the right-hand side by the particle's surface area, $\pi$D\textsuperscript{2}:

\begin{equation} \label{eq5}
\sigma_{ss} = \frac{2.66 \times 10\textsuperscript{-5}}{1+\beta/\alpha}
\end{equation} 

Note that if $\alpha \ll \beta$, $\sigma_{ss} \rightarrow 0$, while if $\alpha \gg \beta$, $\sigma_{ss} \rightarrow 2.66 \times 10\textsuperscript{-5}$ C m\textsuperscript{-2}, the breakdown limit for air at 1 bar.

\begin{figure}[h]
	\centering
	\includegraphics[width=\textwidth]{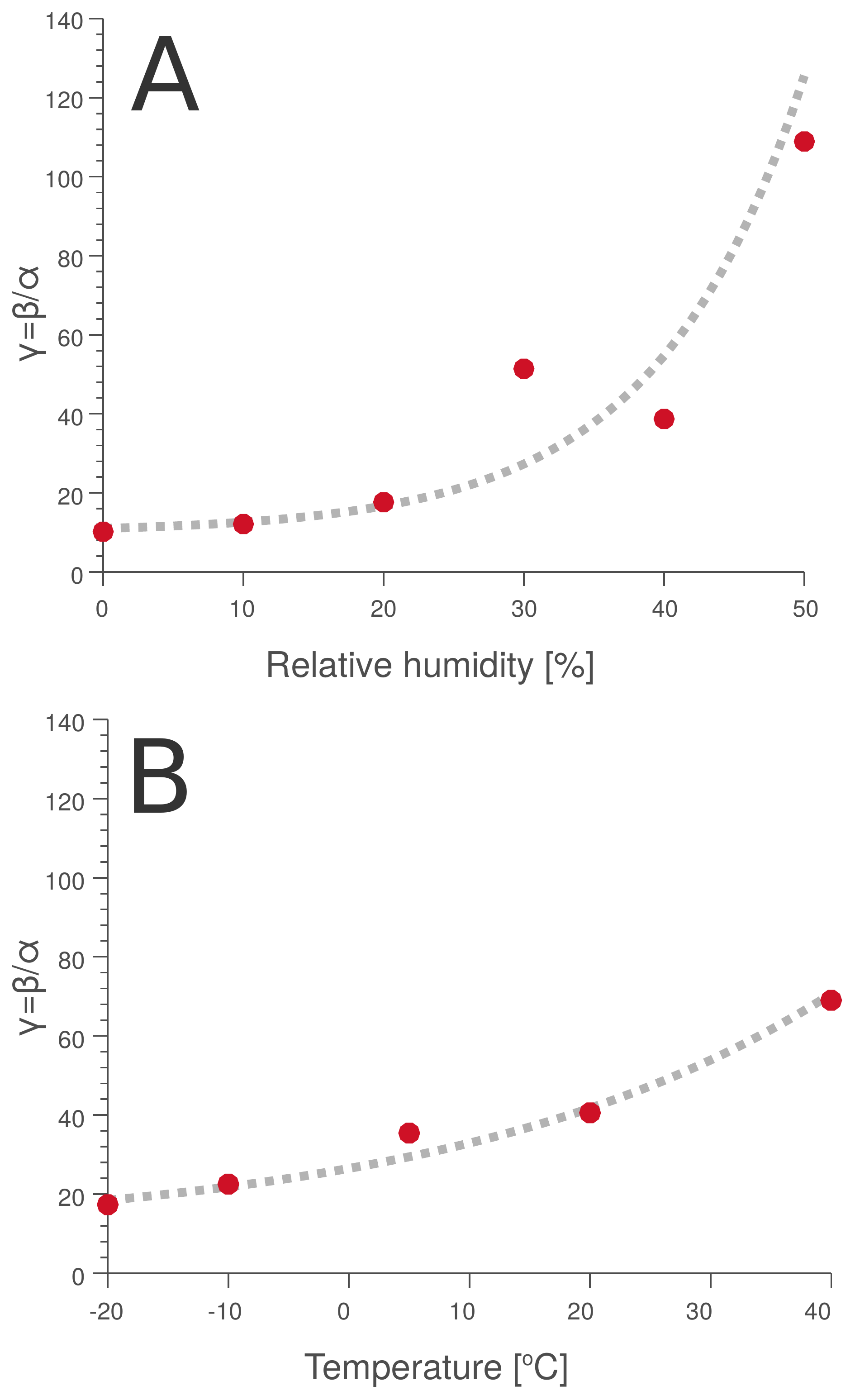}
	\caption{A. The variation of the dimensionless parameter $\gamma$ increases with increasing relative humidity. B. The variation of $\gamma$ increases with increasing temperature.}
	\label{figure4}
\end{figure}

Given our current understanding of triboelectric processes it is difficult to determine whether the RH- and temperature-dependent behaviors we see in our experiments result from a reduced triboelectric charging efficiency (that is, a smaller $\alpha$), an increase in charge recombination processes (larger $\beta$), or both. However, we can explore an overall trend defining the dimensionless parameter $\gamma = \beta/\alpha$. The variation of $\gamma$ for both RH and temperature is rendered in figure \ref{figure4}a and\ref{figure4}b, respectively. We note that the variation of $\gamma$ can be described by an exponential of the form:

\begin{equation} \label{eq6}
\gamma = C_1e^{c_2x} + C_3
\end{equation} 

where $x$ is either temperature or RH and C\textsubscript{1} through C\textsubscript{3} are constants. This equation is plotted as dotted curves in figures \ref{figure4}a and \ref{figure4}b. As can be seen, $\gamma$ is always larger than 1 for the conditions employed in our experiments, indicating that charge-inhibiting mechanisms dominate over processes of charge accumulation. 

The trends in our data may be explained by an increase in particle conductivity brought on by humidity and temperature. \citet{zheng2014theoretical} indicate that humidity controls the stability of charge on particles through the deposition of water films. In other words, humidity modulates the \textit{surface} conductivity of ash grains. Because natural silicate grains tend to have rough surfaces, adsorbed water molecules only form disconnected films that fill pores at low relative humidity. At some critical relative humidity, however, these films connect into a network across a particle's surface, causing a precipitous increase in the surface conductivity. Previous work suggests that this transition occurs around 30-40$\%$ relative humidity. \citet{zheng2014theoretical} propose that increased surface conductivity changes the energy barrier an electron must overcome in order to transfer from one surface to another during a particle-particle collision (in the context of the aforementioned trapped-electron model). Those authors present a numerical model demonstrating that electron tunneling between surfaces is essentially cut off at the critical relative humidity. The steep drop in charge density observed in our experiments between 20 and 30$\%$ relative humidity hints that volcanic materials may follow this trend. Relative humidity may also change the rate at which particles lose charge. While experimental data is sparse, \citet{tada1995direct} provide evidence that humid air may promote direct charge transfer to the atmosphere. However, future work needs to be performed to elucidate the microphysical mechanisms underlying this process. 

As mentioned above, \citet{stern_temp_2019} also investigated the effect of water on the production of charge in a simulated volcanic jet (using a shock tube). It is interesting to note that those experiments required several orders of magnitude more water ($\sim$ 15 wt. \%)  than ours ($\sim$ 10\textsuperscript{-5} wt.\% water) to cause charge production to drop by an order of magnitude. We surmise that these discrepancies arise from differences in granular dynamics, specifically collisional energy, and the timescales between the experiments. Firstly, as shown in \citet{mendez2017thesis}, the shock tube produces high-energy, disruptive particle-particle collisions. Thus, the experiments of \citet{stern_temp_2019} involve fragmentation charging in addition to any frictional electrification. Secondly, newly-formed surfaces may be nominally dry and some time is required for water molecules to absorb to them and create the networks discussed above. Indeed, \citet{telling2013ash} report that ash particles from Tungurahua and Mt. Saint Helens necessitate residence times of several minutes in humid environments for water layers to form on their surfaces (even under humidities approaching 100\%). Because the shock tube decompression event (including electrical signals) has time scales of $\sim$ 10 ms and involves much higher collisional frequencies, we hypothesize that the rate of charge accumulation (both from fragmentation and triboelectric charging) is much greater than the rate of charge loss in that system (i.e. $\alpha >> \beta$). Conversely, our experiments are characterized by periods (tens of minutes) comparable to the absorption timescales reported by \citet{telling2013ash} and the tumbling did not involve disruptive contacts. Together, these behaviors suggest that the accumulation of charge in a mobilized granular flow can be severely hindered by even a small amount of water if residence times of grains are high and collisions produce no new surfaces (i.e. are non-disruptive). 

Unlike humidity, temperature may control both the surface and volumetric electrical conductivities of materials. As discussed above, temperature changes can modify the reactivity of contaminants on particle surfaces. However, because we carefully cleaned the grains before conducting our experiments, we believe that changes in surface conductivities were minimal. Thus, the decrease in charging efficiency with temperature could indicate an increase in the bulk conductivity of the particles. Such effect has been observed in other silicate materials (especially those containing K, Ca, and Na) and has been cited as evidence for an ionic conduction mechanism in glasses  \citep{eldin1998electrical, sasek1981electrical}. These works indicate that increasing temperature can reduce the energy barriers that cations (positive ions) must overcome to migrate within the glass network. The temperature-dependent volumetric conductivity $\rho$ (often expressed in units of $\Omega \cdot$m or $\Omega \cdot$cm) of oxide glasses has been described as:

\begin{equation} \label{eq7}
\log \rho = A + B \frac{1}{T},
\end{equation}

where A and B are composition-dependent coefficients and T is temperature in K. In an excited granular media, the enhanced mobility of positive ions within the bulk of grains could facilitate the recombination of electrons transferred during particle-particle collisions. Presumably, such process would change how charge is partitioned and collected during surface contacts (i.e. modulating the value of $\alpha$). However, \citet{alvarez1978electrical} studied the electrical conductivity of volcanic materials across a wide temperature range and found that increases in conductivity happen at much higher temperatures than those studied here. Those results indicate that the mechanisms operating in synthetic solids may not be applicable to natural volcanic materials. More recently, \citet{woods2020influence} studied the conductivity of a number of ash samples, but did not explore temperature dependence. 

Another possibility is that the decrease in charging we observe with increasing temperature does not reflect a temperature-dependent process but rather an indirect effect of water. While the relative humidity was maintained constant throughout the experiments with variable temperature, the \textit{total water content} in the atmosphere surrounding the colliding particles increased with increasing temperature (see upper axis of figure \ref{figure3}b). Relative humidity, not total water content, is often reported in works dealing with granular electrification because RH controls the growth of water layers on surfaces (i.e. wetting). However, \citet{ogino2019triboelectric} has found that total water content better explains the triboelectric behaviors of polytetrafluoroethylene than relative humidity, suggesting that the growth and distribution of water layers on surfaces are not the only water-based processes modulating the stability of electrical charge. In any case, our results demonstrate that more detailed work on the effect of temperature on electrostatic processes is required.

\section{Insights into the electrification mechanisms operating in volcanic systems}

Both the available water and temperature change in complex manners throughout an eruptive column. Previous work indicates that pre-eruptive magmas typically contain 1-15 weight percent (wt. \%) water \citep{carey1995intensity, grove2005magnesian, ushioda2014water}. Within the conduit and at the vent, elevated temperatures and pressures maintain this magmatic water in solution or in vapor form. Once ejected into the atmosphere, however, the column cools rapidly, causing the relative humidity to increase toward saturation. Atmospheric water may also be entrained into the flow if the column rises through a moist environment. Lastly, water can be added to the system if an erupting volcano is capped with glaciers or if explosions are phreatomagmatic. 
 
 Copious field evidence exists to suggest that at elevation the conversion of this water into ice in fully developed plumes precipitates large scale (several km in length) discharges similar to those in thunderclouds. Indeed, data from Augustine, Redoubt, Eyjafjallaj\"okull, Cord\'on Caulle, and Calbuco reveal that some of the strongest electrical activity took place once the eruptive column climbed to elevations where ice-graupel interactions could occur \citep{thomas_electrical_2007, bennett_monitoring_2010, arason2011charge, nicora2013actividad, vaneaton2016volcanic}. 
 
 In the absence of ice, however, our experimental data suggests that only a small amount of water is required to significantly reduce in the efficacy of triboelectric charging. Indeed, 0 - 0.012 kg m\textsuperscript{-3}, corresponding to an increase in RH from 0-50\% at 25 \textsuperscript{o}C or 0 - 10\textsuperscript{-5} wt.\% water, can hinder triboelectrification in an agitated sample of volcanic ash by up to an order of magnitude (as measured in terms of charge density). Changing temperature appears to also complicate electrostatic effects, but the relationships are less clear--possibly related to the total amount of water in the atmosphere rather than the relative humidity. Yet, our analysis suggests that for small amounts of water to impair tribocharging in plumes, the residence time of ash in a humid environment must be large and collision rates low and non-disruptive. Thus, triboelectrification may be a poor candidate for electrification once pyroclasts have left the energetic gas-thrust region.

 \begin{figure}[h]
	\centering
	\includegraphics[width=\textwidth]{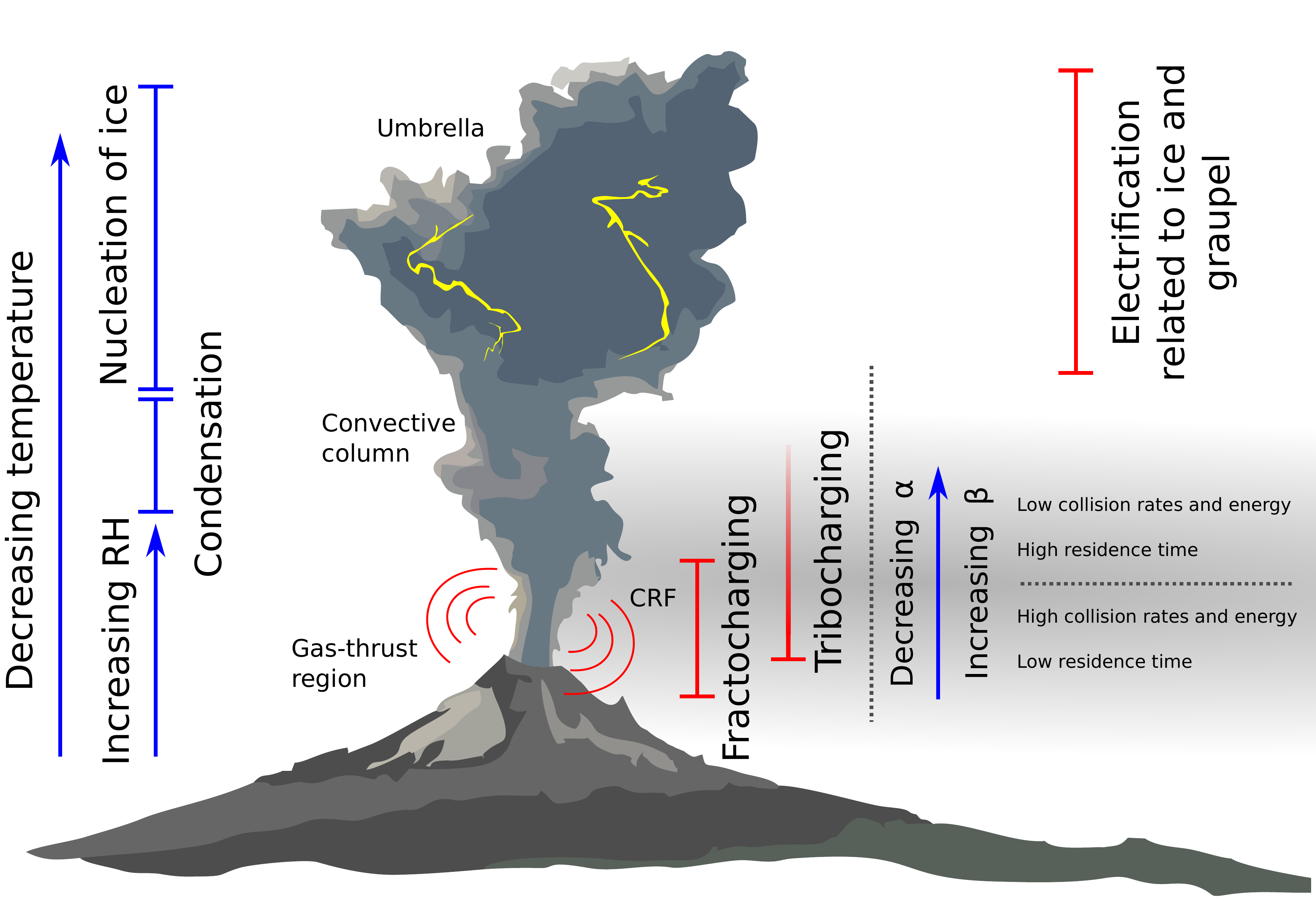}
	\caption{Schematic summarizing charging conditions in a plume in addition to proposed location of triboelectric charging in context with other putative electrification mechanisms. Given the large amounts of water often involved in eruptions, triboelectric charging may be only efficient in the gas-thrust region where collision rates are high and disruptive and residence times are low. Higher in the column, increasing humidity, lower collision rates, and longer residence times may quench electrification from triboelectric process. Electric phenomena may be the result of other mechanisms such as electrification from the interaction of ice and graupel.}
	\label{figure5}
\end{figure}
 
 Alternatively, when collision rates are high and/or produce fragmentation, charging may be less sensitive to water. Field studies, for example, reveal that explosive forcing at the vent is often accompanied by innumerable small discharges that generate powerful continual radio frequency (CRF) emissions. These discharges are typically much smaller in length (on the scale of a few tens to hundreds of meters) than those in mature plumes and occur in the vicinity of the vent \citep{thomas_electrical_2007, aizawa_physical_2016, mendez2018inferring}. During the Augustine (2006) and Redoubt (2009) eruptions, the onset of CRF emissions coincided with explosive activity and the signals had total durations of 2-3 minutes. We note that this vigorous electrical activity at the vent occurs in the absence of ice. Similar to the processes in shock tube experiments (\citet{cimarelli_experimental_2014, mendez2018inferring, stern_temp_2019}), charging at the vent is likely controlled in large part by fractoelectrification as the the magma column fragments and resultant particles undergo disruptive particle-particle collisions (whether or not fractocharging is affected by water should be the focus of future studies). Additionally, because the collisional periods in the gas-thrust region are much shorter than the rate of water film growth, new surfaces created through disruptive interactions may gain charge triboelectrically in nominally higher water contents than those explored in our experiments. Such hypothesis is consistent with the fact that energetic shock tube experiments require more water to produce comparable reductions in charge generation \citep{stern_temp_2019}.
 
 Once the jet loses kinetic energy and transitions into a buoyantly rising column with fewer disruptive interactions, water films will have time to consolidate networks on grain surfaces and reduce the efficacy of triboelectric charging. The dissipation of charge by water on longer timescales may be related to the hiatuses in electrical activity following the explosive phases at Augustine (2006) and Redoubt (2009). Once the column climbs to elevations where water can freeze, vigorous electrical activity may resume as electrification mechanisms similar to those in thunderclouds become active. We render these processes schematically in figure \ref{figure5}
 
\section{Conclusions}

In this work, we have described a set of experiments designed to explore the link between environmental conditions and the triboelectrification of volcanic ash on the scale of individual grains. Our work shows that the charge per unit area gained by ash particles during collisional and frictional interactions decreases with both increasing total water content and increasing temperature. We found that even small amounts of water can diminish the ability of particles to charge triboelectrically if the residence time of the ash in a humid environment is high and collisions non-disruptive. The effect of temperature under similar timescales was less clear and needs to be explored further. We hypothesize that the effect of variable temperature at a constant relative humidity may actually reflect changes in the total water content of the gas. Because the formation of dissipate water films on grain surfaces requires timescales of minutes, triboelectric charging may still be quite important in volcanic systems near the vent where collision rates are high and new surfaces are being produced. In summary, the importance of silicate-silicate tribocharging in the electrification of granular volcanic flows may be limited to the proximal regions of the vent. Other electrification mechanisms, such as those involving ice and graupel, may play more important roles than tribocharging at elevation. 

\acknowledgments
Data availability statement: All data for this paper are contained in the main text
of the present work (figures). Additional detailed description of experimental apparatus found at the Georgia Tech Library Thesis Repository: https://smartech.gatech.edu/handle/1853/58760.

The authors would like to acknowledge (1) the National Science Foundation Graduate
Research Fellowship Program as well as the Blue Waters Graduate Fellowship Program
for providing support for J.M.H. (2) the National Science Foundation Postdoctoral Fellowship Program for L.C. and (3) grant EAR 1645057 for J.D.

\end{document}